# A Scenario for a Natural Origin of Our Universe

Victor J. Stenger
(Dept. of Philosophy, University of Colorado, Boulder CO)
(Dept. of Physics and Astronomy, University of Hawaii, Honolulu HI)

October 16, 2007

A mathematical model of the natural origin of our universe is presented. The model is based only on well-established physics. No claim is made that this model uniquely represents exactly how the universe came about. But the viability of a single model serves to refute any assertions that the universe cannot have come about by natural means.

It is commonly believed that the universe cannot have come about naturally. Although many authors writing at both the popular and academic levels have described various scenarios for a natural origin, usually based on the vague notion of "quantum fluctuations," even they admit that their ideas are speculative and surrender to the prevailing wisdom that the origin of our universe remains unexplained.[1]

What is it that science does when it "explains" some phenomenon? At least in the case of the physical sciences, it builds a mathematical model to describe the empirical data associated with the phenomenon. When that model works well in fitting the data, has passed a number risky tests that might have falsified it, and is at least not inconsistent with other established knowledge, then it can be said to successfully explain the phenomenon. Further discussion on what the model implies about "truth" or "ultimate reality" falls into the area of metaphysics rather than physics, since there is nothing further the scientist can say based on the data. What is more, nothing further is needed for any practical applications. For example, not knowing whether or not electromagnetic fields are real does not prevent us from utilizing the theory of electromagnetic fields.

We have no direct observations of the event we identify as the origin of our universe, "our universe" being the one we live in but with the far greater portion that arose from the same source now out of sight beyond our horizon. This has led some to insist that, as a consequence, science can say nothing about the origin. Here they parrot the familiar creationist argument that because we did not observe humans evolving we can't say anything about human evolution.

But, of course science can and does talk about unobserved and unobservable events. We have plenty of indirect data by which we can test human evolution. We can reconstruct crimes from forensic evidence. In the case of the universe, the inflationary big-bang model nicely accounts for a wide range of increasingly precise observations that bear on the nature of the early universe.[2] When we extrapolate the model back to the earliest moments, we find that the picture becomes surprisingly simple. Assuming the universe came from nothing, it is empty to begin with so we do not have to deal with the complications of matter and its interactions of which we have only incomplete understanding. In fact, to understand an empty universe we need only apply general relativity and quantum mechanics, theories that have enjoyed almost a century of empirical corroboration.

What I will show is that a mathematical model of the origin of our universe based on no more than these well-established theories can be precisely specified. This model is essentially the "no boundary" model proposed over twenty years ago by Hartle and Hawking.[3] I will present a simplified version, following closely the development of Atkatz in his excellent pedagogical review of quantum cosmology.[4] More details are given in the mathematical supplements of my 2006 book, *The Comprehensible Cosmos*, which can also be referred to for a layperson's discussion.[5] The mathematics used is about at the level of a bachelor's degree in physics from an American university.

No claim will made that the model I will describe is actually how our universe actually came about. The model contains no proof of uniqueness. The purpose of this essay is simply to show explicitly that at least one scenario exists for a perfectly natural, non-miraculous origin of our universe based on our best scientific knowledge. In other words, science has at least one viable explanation for the wholly natural origin of our universe, thus refuting any claim that a supernatural creation was required.

## 1. The Wave Function of the Universe

Although we do not yet have a complete theory that unites quantum mechanics and general relativity, nothing prevents us from attempting to combine the two. Consider the Lagrangian

$$L = \frac{3\pi}{4G}\left[-\left(\frac{da}{dt}\right)^2 a + a\left(1 - \frac{a^2}{a_o^2}\right)\right] \qquad (1)$$

where $a$ is some coordinate in one dimension and $G$ is Newton's gravitational constant. This looks rather arbitrary, but is chosen in hindsight. The canonical momentum conjugate to $a$ is

$$p = \frac{\partial L}{\partial\left(\frac{da}{dt}\right)} = -\frac{3\pi}{2G} a \frac{da}{dt} \qquad (2)$$

The Hamiltonian then is

$$\begin{aligned} H &= p\frac{da}{dt} - L \\ &= -\frac{3\pi}{4G} a\left[1 + \left(\frac{da}{dt}\right)^2 - \frac{a^2}{a_o^2}\right] \end{aligned} \qquad (3)$$

Consider the case where $H = 0$. Then

$$1 + \left(\frac{da}{dt}\right)^2 - \frac{a^2}{a_o^2} = 0 \qquad (4)$$

This is precisely one of the Friedmann equations of cosmology that can be derived from Einstein's theory of general relativity:

$$\left(\frac{da}{dt}\right)^2 - \frac{8\pi\rho G}{3}a^2 = -k \tag{5}$$

where $a$ is the radial scale factor of the universe, $a_o^2 = \frac{3}{8\pi\rho G}$, $\rho$ is the energy density ($c = 1$), and $k$ is the parameter indicating whether the universe is open ($k = -1$), critical ($k = 0$), or closed ($k = 1$). Here $k = 1$, indicating a closed universe.

If the universe is empty, then $\rho = \frac{\Lambda}{8\pi G}$, where $\Lambda$ is the cosmological constant, and thus $a_o^2 = \frac{3}{\Lambda}$ is constant. In this case, the scale factor varies with time as

$$a(t) = a_o \cosh\left(\frac{t}{a_o}\right) \quad a > a_o \tag{6}$$

where the origin of our universe is at $t = 0$. Nothing forbids us from applying this for earlier times. For $t < 0$ the universe contracts, reaching $a_o$ at $t = 0$. Thereafter it expands very rapidly providing for what is called *inflation* (see Fig. 1). Note that the scale factor is never smaller than $a_o$.

While not precisely the exponential usually associated with inflationary cosmology, which occurs when $k = 0$, it amounts to the same thing. The fact that $k = 1$ may seem to contradict the familiar prediction of inflationary cosmology that $k = 0$, that is, that the universe is flat. Actually, a highly flat universe is still possible with $k = 1$. Such a universe is simply very large, like a giant inflated balloon, with the part within our horizon simply a tiny patch on the surface. It eventually collapses in a "big crunch" but so far in the future that we can hardly speculate about it except to conjecture that maximum entropy or "heat death" is likely to happen before the collapse is complete. Models for a closed universe exist that are consistent with all current data.[6] In fact, now that we know the universal expansion is accelerating, such models are even more viable.

Now, let us write the Friedmann equation

$$p^2 + \left(\frac{3\pi}{2G}\right)^2 a^2 \left(1 - \frac{a^2}{a_o^2}\right) = 0 \tag{7}$$

This is a classical physics equation. We can go from classical physics to quantum physics is by means of *canonical quantization* in which we replace the momentum $p$ by an operator:

$$p = -i\frac{d}{da} \tag{8}$$

($\hbar = 1$) and introduce a wave function $\psi$, where $H\psi = E\psi$, and $E = 0$.

$$\left[\frac{d^2}{da^2} - \left(\frac{3\pi}{2G}\right)^2 a^2\left(1 - \frac{a^2}{a_o^2}\right)\right]\psi = 0 \tag{9}$$

This is a simplified form of the *Wheeler-DeWitt equation*, where $\psi$ is grandly referred to as the *wave function of the universe*.[7] Using the conventional statistical interpretation of the wave function, $|\psi(a)|^2 da$ represents the probability of finding a universe in an ensemble of universes with *a* in the range *a* to *a + da*.

In general, the wave function of the universe described by the Wheeler-DeWitt equation is a function of functions (that is, it is a *functional*) that specifies the geometry of the universe at every point in 3-dimensional space, with time puzzlingly absent as a parameter. Here we consider the situation where the geometry can be described by a single parameter, the radial scale factor *a*. This should be a reasonable approximation for the early universe and greatly simplifies the problem. Note that zero energy is consistent with the universe coming from "nothing," which presumably has zero energy, without violating energy conservation. In fact, current cosmological observations indicate that the average density of matter and energy in the universe is equal, within measurement errors, to the critical density for which the total energy of the universe was exactly zero at its beginning.

## 2. Quantum Tunneling

Mathematically speaking, the Wheeler-DeWitt equation in this case is simply the time-independent Schrödinger equation for a non-relativistic particle of mass *m* = 1/2 (in units inverse to the units of *a*) and zero total mechanical energy, moving in one dimension with coordinate *a* in a potential

$$V(a) = \left(\frac{3\pi}{2G}\right)^2 a^2 \left(1 - \frac{a^2}{a_o^2}\right) \tag{10}$$

This potential is shown in Fig. 2. Note that it includes the region $a < a_o$.

The mathematical solution of the Friedmann equation for this region is

$$a(\tau) = a_o \cos\left(\frac{\tau}{a_o}\right) \quad 0 < a < a_o \tag{11}$$

where $\tau$ is a real number and the "time" $t = i\tau$ is an imaginary number. Thus the region $a < a_o$ cannot be described in terms of the familiar operational time, which is a real number read off a clock. This is an "unphysical" region, meaning it is a region not amenable to observation. However, meaningful results can still be obtained when our equations are extended into the physical region.

This is a common quantum mechanical procedure, as in the familiar problem of tunneling through a square barrier (see Fig. 3). Inside the barrier the wave function is

$$\psi(x,t) = C \exp[i(px - Et)] + D \exp[i(-px - Et)] \quad -b/2 < x < b/2 \tag{12}$$

where $p = i[2m(V-E)]^{1/2}$ is an imaginary number. Using imaginary time is analogous to using imaginary momentum in this case.

The probability for tunneling through a square barrier, found in textbooks, is

$$P = \exp\left\{-2[2m(V-E)]^{1/2} b\right\} \tag{13}$$

The potential in the Wheeler-DeWitt equation resembles the kind of potential barriers encountered, for example, in nuclear physics where $\alpha$-decay is explained as a quantum tunneling process. A variable barrier $V(x)$ in the range $x_1 < x < x_2$ can be treated as a sequence of infinitesimal square barriers of the type above. The probability for transmission through the barrier will then be,

$$P \approx \exp\left(-2\int_{x_1}^{x_2} dx\, 2m|V(x) - E|^{1/2}\right) \qquad (14)$$

Here, $2m = 1$, $E = 0$, $x_1 = 0$, and $x_2 = b = a_o$, so

$$P \approx \exp\left(-2\int_{o}^{a_o} da\, K(a)\right) \qquad (15)$$

where

$$K(a) = \left(\frac{3\pi}{2G}\right)^2 a \left(\frac{a^2}{a_o^2} - 1\right)^{1/2} \qquad (16)$$

The above integral can be shown to yield,

$$P \approx \exp\left(-\frac{3}{8G^2\rho}\right) \qquad (17)$$

At the earliest definable moment, the Planck time ($1.61\times10^{-43}$ second), we estimate from the uncertainty principle that $\rho$ will be on the order of the Planck density, $\rho = 1/G^2 = 3\times10^{126}$ electron-volts per cubic centimeter. In that case, the tunneling probability is exp(-3/8) = 68.7 percent. This suggests that the unphysical region is unstable and that 68.7 percent of all universes will be found in the physical state.

The region $a < a_o$ is a classically disallowed region that is allowed by quantum mechanics and described by a solution of the Wheeler-DeWitt equation. The particular solution will depend on boundary conditions. In their no-boundary model, Hartle and Hawking propose equal amounts of incoming and outgoing waves, which Atkatz writes

$$\psi_{HH}(a > a_o) = K(a)^{-1/2} \cos\left[\frac{\pi}{2}a_o^2\left(\frac{a^2}{a_o^2} - 1\right)^{3/2}\right] \qquad (18)$$

$$\psi_{HH}(0 < a < a_o) = |K(a)|^{-1/2} \exp\left[-\frac{\pi}{2}a_o^2\left(1 - \frac{a^2}{a_o^2}\right)^{3/2}\right] \qquad (19)$$

These solutions have been obtained using WKB approximation and do not apply for the regions around $a = 0$ and $a = a_o$. The wave function is shown in Fig. 4, where I have extrapolated to zero at $a = 0$ and smoothly connected the inside and outside solutions at $a = a_o$.

## 3. Interpretation

The simplified Hartle-Hawking model gives one possible scenario for the universe to come about naturally. In this picture, another universe existed prior to ours that tunneled through the unphysical region around $t = 0$ to become our universe. Critics will argue that we have no way of observing such an earlier universe, and so this is not very scientific. However, the model is based on well-established theories that make no distinction between the two sides of the time axis. Nothing in our knowledge of physics and cosmology requires the non-existence of that prior universe, so it would be a violation of Occam's razor to exclude it.

However, if you insist on no prior universe you can use an alternate solution to the same Schrödinger equation provided by Vilenkin where the universe simply starts at $t = 0$. It is just a matter of setting a boundary condition.[8] Vilenkin interprets his solutions as tunneling from "nothing," where he takes the classically unphysical region $a < a_o$ to represent nothing.

While not crucial to our scenario of a natural origin for our universe, it is interesting to note that the earlier universe in the Hartle-Hawking case is only earlier from our point of view, where the arrow of time points away from the big bang in the direction of increased entropy. Our sister universe will have an arrow of time pointing opposite to ours since that is the direction of increased entropy on the negative side of the time axis. The two universes, then, are mirror images of one another, both emerging simultaneously from the same point in space and time. They are not likely to be identical, however, since we expect from our current models of cosmology and particle physics that random processes play a major role in the development of structure.

This scenario then provides another way of looking at the origin of the universe. Two universes "begin" in the unphysical region around $t = 0$ that Vilenkin calls "nothing." These universes then evolve along the opposite directions of the time axis. In either case, the basic mechanism is quantum tunneling, as established phenomenon that is less vague than "quantum fluctuation."

## Notes

[1] The original idea is usually attributed to E. P. Tryon, "Is the universe a quantum fluctuation?" *Nature* 246 (1973): 396-97.

[2] Alan Guth. "The Inflationary Universe: A Possible Solution to the Horizon and Flatness Problems," *Physical Review* D23 (1981): 347-56; Alan Guth. *The Inflationary Universe*, New York: Addison-Wesley, 1997; Linde, André. "A New Inflationary Universe Scenario: A Possible Solution of the Horizon, Flatness, Homogeneity, Isotropy, and Primordial Monopole Problems." *Physics Letters 108B* (1982): 389-92; "Quantum Creation of the Inflationary Universe," *Lettere al Nuovo Cimento* 39 (1984): 401-05.

[3] James B. Hartle and Stephen W. Hawking, "Wave Function of the Universe," *Physical Review* D28 (1983): 2960-75.

[4] David Atkatz, "Quantum cosmology for pedestrians," *American Journal of Physics* 62 (1994): 619-27.

[5] Victor J. Stenger, *The Comprehensible Cosmos: Where Do The Laws Of Physics Come From?* (Amherst, NY, Prometheus Books, 2006).

[6] Marc Kamionkowski and Nicolaos Toumbas, "A Low-Density Closed Universe, *Physical Review Letters* 77 (1996): 587-90.

[7] B.S. DeWitt, "Quantum Theory of Gravity. I. The Canonical Theory," *Physical Review* 160 (1967): 1113-48, J.A. Wheeler, "Superspace and the nature of quantum geometrodynamics," in C. DeWitt, and J.A. Wheeler, eds., *Battelle Rencontres*: *1967 Lectures in Mathematics and Physics*, (New York:W.A. Benjamin, 1968).

---

[8] Alexander Vilenkin, "Boundary conditions in quantum cosmology," *Physical Review D* 33 (1986): 3560-69. The Vilenkin wave function is given in *The Comprehensible Cosmos*.

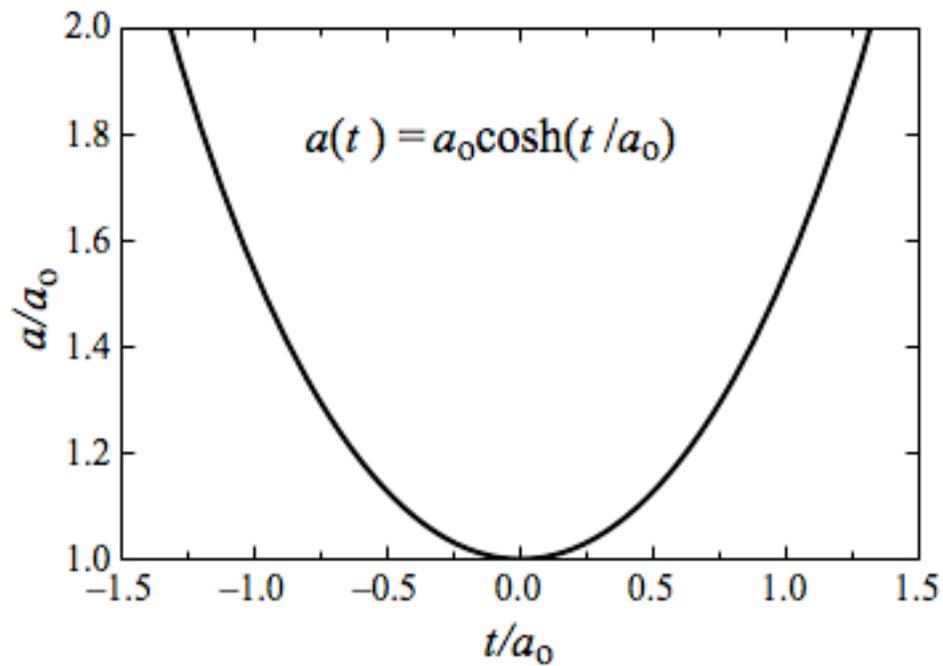

Fig. 1. The solution of Friedmann's equation for an empty universe with $k = 1$. The region $t < 0$ is not forbidden by any known principle.

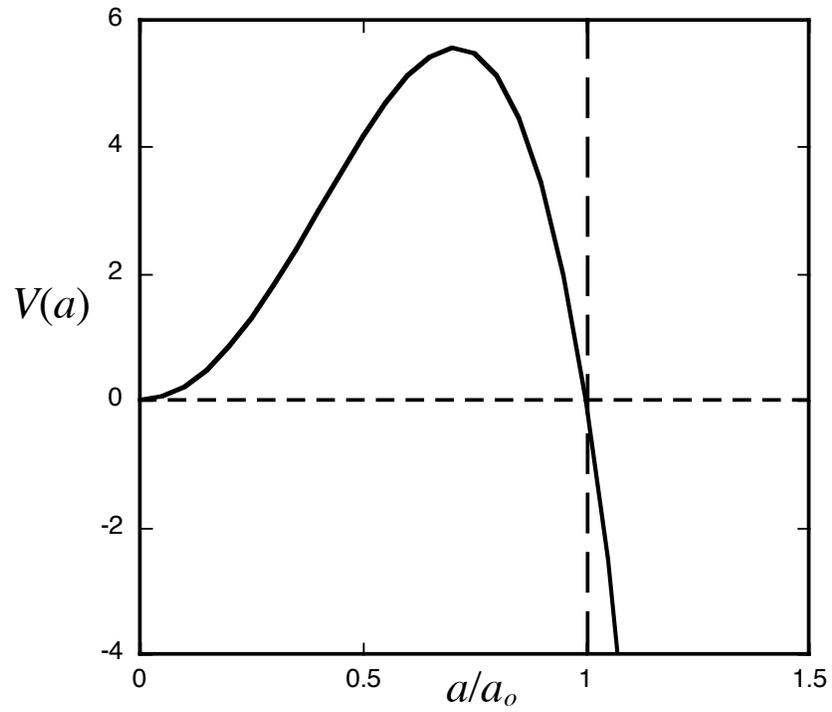

Fig. 2. The potential energy in the Wheeler-DeWitt equation. The vertical scale is in units of $\left(\dfrac{3\pi}{2G}\right)^2$.

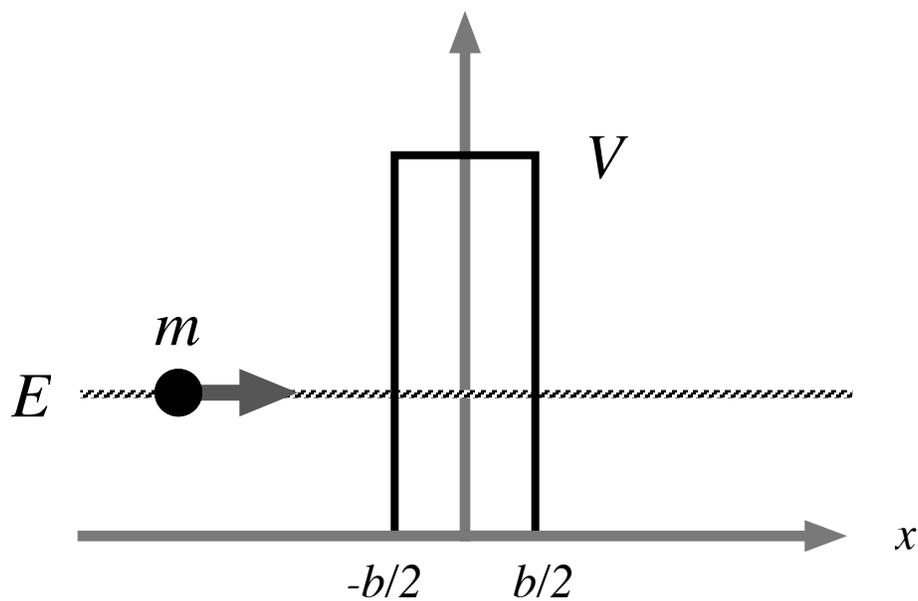

Fig. 3. A particle of mass $m$ and energy $E$ incident on a potential energy barrier of height $V$ and width $b$. Quantum mechanics allows for particles to tunnel through the barrier with a calculable probability. The region inside the barrier is "unphysical," that is, unmeasurable, with the particle's momentum being an imaginary number.

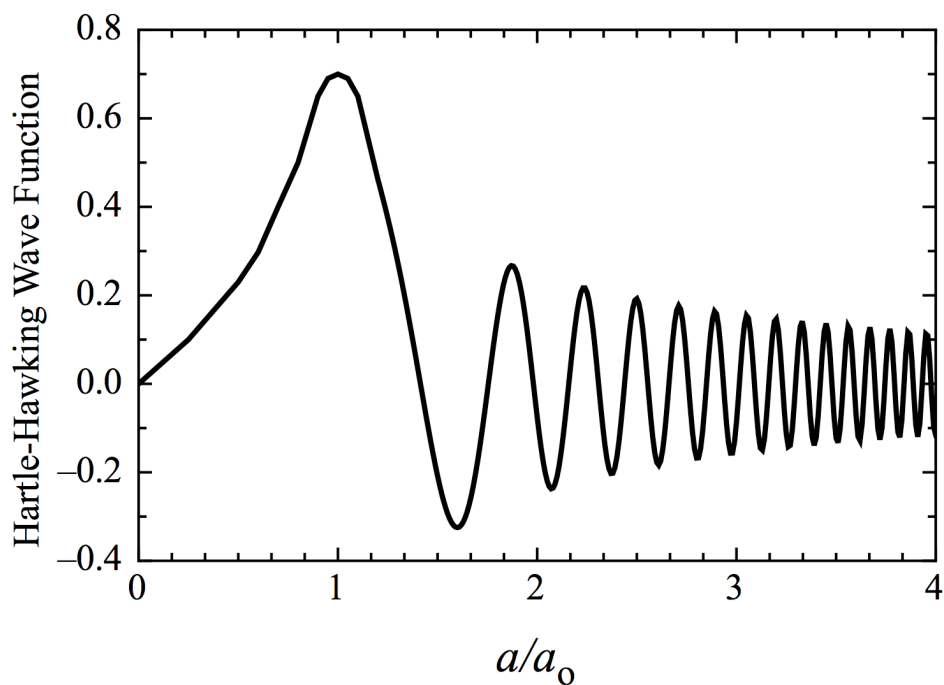

Fig. 4. The Hartle-Hawking wave function of the universe.